\documentclass[12pt]{article}
\usepackage{amssymb}
\usepackage{amsmath}
\usepackage{latexsym}
\usepackage{epsfig}
\usepackage{graphicx}
\usepackage{subfigure}
\usepackage{wasysym}
\usepackage{threeparttable}
\usepackage[authoryear,comma,sort&compress]{natbib}
\usepackage{color}
\newcommand{\mbv}[1]{\mbox{\boldmath$#1$\unboldmath}}
\newcommand{\mbf}[1]{\mathbf{#1}}

\usepackage{hyperref}

\def\diag{\hbox{diag}}

\newcommand{\Appendix}
{
\def\thesection{Appendix~\Alph{section}}
\def\thesubsection{A.\arabic{subsection}}
}
\def\diag{\hbox{diag}}

\def\boxit#1{\vbox{\hrule\hbox{\vrule\kern6pt
          \vbox{\kern6pt#1\kern6pt}\kern6pt\vrule}\hrule}}

\def\bse{\begin{eqnarray*}}
\def\ese{\end{eqnarray*}}
\def\be{\begin{eqnarray}}
\def\ee{\end{eqnarray}}
\def\bq{\begin{equation}}
\def\eq{\end{equation}}
\def\bse{\begin{eqnarray*}}
\def\ese{\end{eqnarray*}}

\begin{document}
\thispagestyle{empty} \baselineskip=28pt

\begin{center}

{\LARGE{\bf Spatial Fay-Herriot Models for Small Area Estimation with Functional Covariates}}
\end{center}

\baselineskip=12pt

\vskip 2mm
\begin{center}
Aaron T. Porter\footnote{(\baselineskip=10pt to whom correspondence should be addressed) Department of Statistics, University of Missouri-Columbia,
146 Middlebush Hall, Columbia, MO 65211, porterat@missouri.edu}, Scott H. Holan\footnote{\baselineskip=10pt  Department of Statistics, University of Missouri-Columbia, 146 Middlebush Hall, Columbia, MO 65211-6100},
Christopher K. Wikle$^2$, 
Noel Cressie$^{2,\,}$\footnote{\baselineskip=10pt National Institute for Applied Statistics Research Australia (NIASRA), School of Mathematics and Applied Statistics, University of Wollongong, NSW 2522, Australia.}

\end{center}
%
%
%
%
\vskip 4mm

\begin{center}
{\bf \large{Abstract}}
\end{center}
The Fay-Herriot (FH) model is widely used in small area estimation and uses auxiliary information to reduce estimation variance at undersampled locations. We extend the type of covariate information used in the FH model to include functional covariates, such as social-media search loads or remote-sensing images (e.g., in crop-yield surveys). The inclusion of these functional covariates is facilitated through a two-stage dimension-reduction approach that includes a Karhunen-Lo\`{e}ve expansion followed by stochastic search variable selection. Additionally, the importance of modeling spatial autocorrelation has recently been recognized in the FH model; our model utilizes the intrinsic conditional autoregressive class of spatial models in addition to functional covariates.  We demonstrate the effectiveness of our approach through simulation and analysis of  data from the American Community Survey. We use Google Trends searches over time as functional covariates to analyze relative changes in rates of percent household Spanish-speaking in the eastern half of the United States.

\baselineskip=12pt

%
%
%

\baselineskip=12pt
\par\vfill\noindent
\par\medskip\noindent
\clearpage\pagebreak\newpage \pagenumbering{arabic}
\baselineskip=24pt
\section{Introduction}\label{sec:Intro}
The Fay-Herriot (FH) model \citep{fay1979estimates} is one of the primary tools used in small area estimation (SAE) \citep[e.g.,][among others]{roy2007empirical,jiang2011best,you2011hierarchical}. Model-based estimates are widely used in SAE as they represent a way to borrow strength across locations and thereby reduce the mean squared errors (MSE) of the small area estimates \citep{rao2003small}. These models utilize scalar auxiliary information to obtain an ``indirect" estimate of the small-area variable of interest, rather than using a direct survey estimate.

As government budgets remain flat or decline, auxiliary information that is relatively inexpensive and readily available, but that is still representative of the population under consideration, is of substantial interest. Functional covariates based on internet sources, social media, or other sources (e.g., remotely sensed image data) may augment or replace scalar auxiliary information for a wide variety of surveys.  The advantage of these types of covariates is that they are often readily available and provide significant information related to a diverse set of demographic and other survey outcomes. For instance, Twitter tweets or Google searches can be associated with a precise location and searched for specific hashtags or terms.  Further, dimension-reduced representations of satellite imagery could be used as auxiliary information in modeling outcomes from agricultural surveys.

Not surprisingly, many federal agencies (including the United States (U.S.) Bureau of Labor Statistics, U.S. Census Bureau, among  others) have now realized the potential importance of harnessing these massive, readily available data sources.  Methodologies relying on ``web-scraping" for the collection of data and use of retail scanner and social-media data have emerged as avenues of particular interest \citep[e.g.,][]{capps2013roward, horrigan2013big}. Consequently, it is extremely important that sound and effective statistical methodology be developed to accommodate this abundantly rich class of ``Big Data" resources.

Functional data analysis (FDA) methodology allows for the use of curves, images, and other ``objects" as either independent or dependent variables in a statistical modeling framework \citep[e.g.,][among others]{ramsay2005,ramsay2006}. The use of FDA in a (generalized) linear statistical modeling framework is well developed, with a substantial amount of research occurring over the last decade. For example,  \cite{goldsmith2012long} develop scalar-on-functional regression, where it is assumed that the scalar response is a member of the exponential family of distributions. \cite{james2002generalized} considers generalized linear models with both functional covariates and a functional response, and \cite{muller2005generalized} utilize a Karhunen-Lo\`{e}ve expansion for functional covariates when modeling a scalar response. From a Bayesian perspective, \cite{baladandayuthapani2008bayesian} work with spatially correlated functional data, and \cite{crainiceanu2009generalized} develop multilevel functional regression models. Aside from the particular applications being considered, there are several key distinctions between our approach and that of \cite{baladandayuthapani2008bayesian}, including where the spatial-correlation structure is placed and what type of correlation structure is imposed.  In \cite{baladandayuthapani2008bayesian} the spatial correlation is based on a Euclidian distance between functions and given a Mat\'{e}rn structure.  In contrast, our approach conditions on functional covariates and uses an  intrinsic conditional autoregressive (ICAR) spatial structure. Moreover, we find the most predictive functional components using stochastic search variable selection.

Survey sampling followed by SAE is commonly implemented by official-statistics agencies, but in this article we propose a shift from the usual FH model. We propose a spatial FH model that uses functional and/or image covariates as auxiliary information. This innovative combination of incorporating spatial dependence along with functional and/or image covariates simultaneously leverages information from multiple sources to provide more precise small area estimates. Examples of such functional/image covariates include Google Trends curves, Twitter hashtag counts, and remotely sensed satellite imagery. The use of social media and other internet-based predictors is a developing field \citep[see, e.g.,][]{signorini2011use}. However, FH modeling employing such functional data (covariates) and spatial dependence remains undeveloped, and our article addresses SAE using such models.

Within the frequentist setting there have been several attempts at incorporating spatial dependence into the FH model through the use of simultaneous autoregressive (SAR) models \citep[e.g., see][among others]{singhspacetime2005, pratesiSAR2009, molinabootstrap2009}; applications outside of official statistics can also be found \citep{petrucciwatershed2006}. Our approach proceeds from a Bayesian perspective and, thus, it allows a natural quantification of uncertainty through posterior distributions. The expected posterior variance is simply the MSE of the relevant small area estimate \citep[e.g., ][p. 38]{wiklecressie}.  The Bayesian paradigm provides a natural hierarchical framework for incorporating latent spatial random effects. In particular, we propose a FH model that utilizes ICAR random effects to capture spatial dependence. Finally, we use functional covariates that are (dimension-reduced) temporal curves generated from Google Trends \citep{googletrends},  in a statistical model of state-level American Community Survey (ACS) data (\url{http://www.census.gov/acs}).

The ACS is an on-going survey performed by the U.S. Census Bureau that provides single-year and multiyear estimates for a large number of demographic variables. Publicly available data provide one-year estimates for areas with large populations (e.g., locations with over 65,000 individuals),  three-year-period estimates for areas with over 20,000 individuals, and five-year-period estimates for all areas. The public-use microdata samples (PUMS) are also available for a diverse set of variables and can be used to model smaller geographies, known as public use microdata areas (PUMAs) (see\\ \url{http://www.census.gov/acs/www/data_documentation/public_use_microdata_sample/} for comprehensive details). The methodology we present here could also be used to fit statistical models to PUMS.

SAE is typically performed on smaller geographies than states, such as at the county level or the census-tract level. Our reason for analyzing data using each state as a unit is that currently the Google Trends data are available at the state level (although one can also obtain search data for the ten largest cities in any state).  It is important to emphasize that, for any particular problem, it is possible that  other functional/image data (such as Twitter or other social-media data) may be available at smaller geographies, and our methodology is equally applicable in this case.

The structure of this paper is as follows. We first introduce the motivating data in Section~\ref{SecMotive}. We provide the methodological details of our approach in Section~\ref{SecTheory}, and we demonstrate reduction in MSE through a simulation study in Section~\ref{SecSims}.  An analysis using the proposed methodology, in the context of ACS data on the rate of change in percent household Spanish-speaking in the Eastern United States, is given in Section~\ref{SecData}. We close with a discussion in Section~\ref{SecDis}.  For convenience of exposition, relevant computational details can be found in two Appendices.

\section{Motivating Data: The American Community Survey}\label{SecMotive}
The variable ``relative change of percent household Spanish-speaking" in different areas of the U.S. may provide insight into immigration patterns as well as provide a marker for socio-economic factors. The standard errors of the ACS estimates for variables associated with language spoken tend to be larger than most other variables in the survey, and this is even true at larger geographies, such as at the state level. To improve estimates, we incorporate Google Trends data \citep{googletrends} as auxiliary information in a framework that uses a spatial FH model with functional covariates. Google Trends data provide state-level weekly time series indicating scaled search loads in various categories (e.g., see Figure~\ref{Fi:Functional}).

By considering Google Trends searches that contain commonly used Spanish words, we are able to develop a proxy measure for percent household Spanish-speaking. It is reasonable to assume that individuals who speak Spanish at home are more likely to perform internet searches in Spanish. The ubiquitous presence of Google and other social-media services make these searches a readily available source of data.

When determining which Google Trends data should be used as a proxy for the pattern of percent household Spanish-speaking, our approach was to analyze the Google searches of relatively common Spanish words. Several candidate words were selected, and we found relatively high search volume for the words ``y," ``el," and ``yo," which mean ``and," ``the," and ``I" in English, respectively. These words rarely appeared in searches in other languages. We base our simulation study (Section~\ref{SecSims}) and application (Section~\ref{SecData}) on these search results.

Google Trends data present several issues that must be addressed prior to analysis. The first issue is related to the way that Google Trends data are defined.\footnote{\baselineskip=12pt The Google Trends data used in this article were downloaded prior to October 2012.  Subsequently, Google changed the normalization applied to the data and, therefore, the Google Trends data, as presented here, are no longer available for download; however, they are available upon request from the corresponding author. Nevertheless, the methodology presented in this article is equally applicable to the currently available Google Trends data.} Although they can be scaled and normalized to a fixed time point by state, the raw data cannot be directly accessed \citep{googletrends}. This means that the values of the Google Trends data cannot be compared between states, and only within-state comparisons across time are valid. To remedy this problem, we fix the time frame of $2008-2009$ as our period of interest, and we standardize each curve to have a within-curve mean of zero and a within-curve standard deviation of one. This results in curves with the same scale from state to state, which facilitates extraction of curve features, rather than spurious differences in magnitude.

Because we have considered search loads from $2008-2009$, we need to perform some standardization of the outcome. The ACS data that we consider for each state is the relative change of percent household Spanish-speaking, which is defined as
\begin{eqnarray}
\frac{\%\,\, \mbox{household Spanish-speaking in 2009} - \%\,\,\mbox{household Spanish-speaking in 2008}}{ \%\,\, \mbox{household Spanish-speaking in 2008}}.
\label{eqn:Outcome}
\end{eqnarray}
The western and eastern halves of the US may behave differently with regard to relative change of percent household Spanish-speaking; so, for illustration, we restrict our analysis to 20 states and the District of Columbia in the eastern half of the U.S. This yields 21 locations of interest, many of which have traditionally had a low number of native Spanish speakers. As a consequence, relatively large changes may appear in the ACS, but are they real?  The margins of error (MOE) for the ACS estimates of percent household Spanish-speaking tend to be larger in the eastern half of the country. Considering states in the eastern half of the U.S. as small areas gives the FH model the potential to provide a great deal of improvement when compared to the public-use ACS estimates.

Iowa, Mississippi, Arkansas, Virginia, West Virginia, Delaware, Rhode Island, Vermont, New Hampshire, and Maine are excluded from our analysis. There were two reasons that a state was excluded from consideration. The first is that the search load for more than 20\% of the weeks under consideration did not meet the threshold that Google Trends uses to indicate search loads. When the threshold was not met, Google Trends reports the value to be zero.   Removing states with 20\% or more zeroes helped to mitigate Google Trends' censoring of the data. The second reason a state was eliminated was because after January 1, 2010, Google Trends redefined, and presumably improved, their algorithm for tagging searches to a location \citep{googletrends}.   Certain states, such as Virginia, exhibited markedly different behavior after that date, which casts doubt on the accuracy of the search loads during the period $2008-2009$ that we considered. Thus, we excluded these states from our analysis.  The number of states (i.e., small areas) considered in our analysis is $n=21$, and they are listed in Table~\ref{Ta:Sim}.

The approach presented here is certainly not unique to estimating relative change of percent household Spanish-speaking. Internet searches or social-media sources contain high-dimensional data that, in principle, could be used in many applications of SAE, thus increasing the types of auxiliary information that could be used to improve survey-based estimates.

\section{Functional Covariates in the Fay-Herriot Model}\label{SecTheory}
The model we propose can be viewed as an extension of the traditional FH model.  Specifically, we propose including functional covariates as a source of auxiliary information, and we propose a random effect that captures spatial correlation. To model the spatial correlation, we use an ICAR structure.

For $i=1,\ldots,n$, the traditional FH model is given by
\begin{eqnarray}
Y_{i}&=&\theta_{i}+\epsilon_{i},
\label{eqn:Main}\\
 \theta_{i}&=& \beta_0 + \mbf{x}_{i}' \mbv{\beta}_{x}+u_{i},
\label{eqn:FH}
\end{eqnarray}
where $\epsilon_i\sim N(0,\sigma^2_i)$ and $u_i\sim N(0,\sigma_u^2)$, with all error terms, $\{\epsilon_i\}$ and $\{u_i\}$, mutually independent.   Here, $\theta_{i}$ is the superpopulation mean of the parameter of interest for small area $i$ and the quantity we wish to estimate; $Y_{i}$ is a design-unbiased estimate of $\theta_{i}$, and the variance of $\epsilon_{i}$, $\sigma_{i}^{2}$, is estimated based on the survey design and assumed known, for $i=1,\ldots,n$.  The auxiliary information at the $i$-th small area is a $q$-dimensional vector of scalar covariates denoted by $\mbf{x}_{i}$,  with associated $q$-dimensional regression parameters $\mbv{\beta}_{x}$ and intercept $\beta_0$. Note that, in this context, the assumption of known sampling-error variances $\{\sigma_i^2$\} is fairly common \citep[e.g., see][]{rao2003small, wang2011bayesian}.

There is an alternate representation of (\ref{eqn:Main}).  If we let $[A\vert B]$ represent the conditional distribution of the random variable $A$ given the random variable $B$, then (\ref{eqn:Main}) can be written as
\begin{eqnarray*}
[Y_i\vert \theta_i, \sigma^2_i]=(2\pi\sigma_i^2)^{-1/2}\exp\left\{-\frac{1}{2}(Y_i-\theta_i)/\sigma_i^2\right\}.
\end{eqnarray*}  Then, under the assumption of independent $\{\epsilon_i\}$, the distribution $\prod_{i=1}^n[Y_i | \theta_i, \sigma^2_i]$ is the ``data model," following the hierarchical modeling terminology in \cite{wiklecressie}. This representation clarifies that the data responses are specified conditionally on the superpopulation mean and sampling error.

\subsection{Dimension-Reduced Functional Covariates}\label{subsec:dimred}
Let $z_{ij}(t),\, t\in\cal{T}$, denote the $j$-th functional covariate ($j=1,\ldots,J$) associated with the $i$-th small area ($i=1,\ldots,n$) defined over the time domain $\cal T$. Note that one could also include spatially indexed functional covariates or image covariates \citep[e.g.,][]{holan2010modeling,holan2012approach} in this framework.   However, for illustration, we focus here on temporal functional covariates.

An extension of model (\ref{eqn:FH}) that includes $J$ functional covariates, can be written as
\begin{eqnarray}
\theta_{i} &=& \beta_0+ \sum_{j=1}^J\int_{\cal T}  \beta_{j}(t) z_{ij}(t)\,dt+ \mbf{x}_{i}' \mbv{\beta}_{x}+u_{i};\,\,\, i=1,\ldots,n,
\label{eqn:FFH}
\end{eqnarray}
where $\{\beta_{j}(t): t\in\cal{T}\}$ is a square-integrable functional parameter associated with the $j$-th functional covariate.  For ease of exposition, we temporarily assume that $J\equiv 1$ and suppress the subscript $j$.  Now, assume that $\{\phi_{k}(t): k=1,2,\ldots\}$ forms a complete orthonormal basis in $\cal{T}$. Then, we have the unique representation,
\begin{equation}
z_{i}(t)=\sum_{k=1}^\infty \xi_{i}(k)\phi_{k}(t),
\label{eqn:expansion2}
\end{equation}
where $\{\xi_{i}(k): k=1,2,\ldots\}$ are expansion coefficients of $z_{i}(\cdot)$, a functional covariate associated with the $i$-th small area.  We also have the unique representation,
\begin{equation}
\beta(t)=\sum_{k=1}^\infty b(k)\phi_{k}(t),
\label{eqn:betas}
\end{equation}
where $\{b(k): k=1,2,\ldots\}$ are the expansion coefficients of $\beta(\cdot)$, which recall is a square-integrable functional parameter.  From the orthonormality property of the basis functions and upon substitution of (\ref{eqn:expansion2}) and (\ref{eqn:betas}), for $J=1$, the model (\ref{eqn:FFH}) can be alternatively expressed  as
\begin{eqnarray}
\theta_{i} &=& \beta_0+ \sum_{k=1}^\infty b(k)\xi_{i}(k) + \mbf{x}_{i}' \mbv{\beta}_{x}+u_{i}.
\label{eqn:FFHdisc}
\end{eqnarray}
Note that (\ref{eqn:FFHdisc}) is a general model that allows for both functional and scalar covariates to be used simultaneously as auxiliary information.  However, in our simulation study and analysis of ACS's percent household Spanish-speaking data, we only utilize functional covariates.  Finally, the case where $J>1$ follows immediately using an identical functional decomposition.

In principle, any complete orthonormal basis set could be used to represent the functional covariates.  In our analysis, we utilize a Karhunen-Lo\`{e}ve (K-L) expansion; see \citet[][Chapter 12]{jolliffeprincipal}, \citet[][Chapters 4 and 5]{wiklecressie}, and the references therein.  The K-L expansion is a commonly used expansion in spatio-temporal modeling (where the basis functions are often empirical orthogonal functions) and functional data modeling (also referred to as functional principal components).  Due to the fact that the expansion is constrained to be orthogonal, only the first component is typically interpretable. In the context of SAE, this is not a concern, as prediction is usually the primary goal.

We continue with the exposition assuming $J=1$ and suppressing the subscript $j$. Following \citet[][Chapter 5]{wiklecressie}, assume that $\{z_{i}(\cdot)\}$ are stochastic processes with $E(z_{i}(t)) = 0$,  and for $t,t' \in \cal{T}$, define the temporal covariance function for the functional covariate as $C_{0}(t,t') = E(Z_{i}(t) Z_{i}(t'))$, which is assumed to be invariant across small areas (see Cressie and Wikle, 2011, p. 267, for an analogous definition of a spatial covariance function that is invariant in time). Thus, the subscript ``0'' serves to remind us that this is effectively a spatio-temporal covariance function for ``lag 0'' in space and is invariant over all spatial small areas.  Then, assuming this covariance is continuous and square-integrable, we can write
\begin{eqnarray*}
C_{0}(t,t')=\sum_{k=1}^{\infty} \lambda_{k} \psi_{k}(t) \psi_{k}(t'),
\end{eqnarray*}
where $\lambda_{1} \geq \lambda_{2}\geq\cdots \geq 0$ are the eigenvalues and $\{\psi_{k}(\cdot):k=1,2,\ldots\}$ are the orthonormal eigenfunctions that solve the Fredholm integral equation \citep[e.g.,][p. 457-461]{papoulis},
\begin{eqnarray}
\int_{\cal{T}} C_{0}(t,t') \psi_{k}(t')dt'=\lambda_{k} \psi_{k}(t); \,\,\, k=1,2,\ldots,t\in\cal{T}.
\label{eqn:FIE}
\end{eqnarray}
Because the eigenfunctions, $\{\psi_{k}(\cdot): k=1,2,\ldots\}$, form a complete orthonormal basis, $z_{i}(t)$ can be written as,
\begin{eqnarray}
z_{i}(t)=\sum_{k=1}^{\infty} \xi_{i}(k)\psi_{k}(t),
\label{eqn:KLexpansion}
\end{eqnarray}
where $\{\xi_{i}(k): k=1,2,\ldots\}$ are uncorrelated, mean-zero, variance $\{\lambda_{k}: k=1,2,\ldots\}$ random variables, respectively. Thus, one can see that the K-L temporal basis functions $\{\psi_{k}(t)\}$ in (\ref{eqn:KLexpansion}) play the role of the general temporal basis functions $\{\phi_{k}(t)\}$ in (\ref{eqn:expansion2}).

In practice, for $T$ discrete times $\{t_{1},t_{2},\ldots,t_{T}\}$, the {\it empirical} temporal basis functions, $\widetilde{\mbv{\psi}}_{k}\equiv(\widetilde{\psi}_{k}(t_{1}),\ldots,\widetilde{\psi}_{k}(t_{T}))'$, are obtained from a {\it numerical solution} of (\ref{eqn:FIE}). For cases where the discrete times are equally spaced, this is equivalent to solving the spectral decomposition of the empirical temporal covariance matrix \citep[e.g.,][Chapter 5]{wiklecressie}; that is, decompose $\widehat{\mbf{C}}_{0}=\widetilde{\mbv{\Psi}}\widetilde{\mbv{\Lambda}}\widetilde{\mbv{\Psi}}'$, where $\widetilde{\mbv{\Psi}}\equiv \{\widetilde{\mbv{\psi}}_{1},\ldots,\widetilde{\mbv{\psi}}_{T}\}$ is a $T\times T$ matrix, $\widetilde{\mbv{\Lambda}}\equiv \diag(\widetilde{\lambda}_{1},\ldots, \widetilde{\lambda}_{T})$, and $\widehat{\mbf{C}}_{0}\equiv (n-1)^{-1}\sum_{i=1}^{n} (\mbf{z}_{i}-\widehat{\mbv{\mu}})(\mbf{z}_{i}-\widehat{\mbv{\mu}})'$, for $ \widehat{\mbv{\mu}}\equiv n^{-1} \sum_{i=1}^{n} \mbf{z}_{i}$ and $\mbf{z}_{i} \equiv (z_{i}(t_{1}),\ldots,z_{i}(t_{T}))'$. Note, in some applications, one may consider $\widehat{\mbv{\mu}} \equiv \widehat{\mu}_{\cdot } \mbf{1}$, where $\widehat{\mu}_{\cdot }$ is the grand mean, $\widehat{\mu}_{\cdot } \equiv (nT)^{-1}\sum_{i=1}^{n}\sum_{t=1}^{T} z_{i}(t)$, for the functional covariate. A comprehensive discussion of issues associated with the calculation of empirical basis functions in the discrete K-L framework can be found in \citet[][Chapter 5]{wiklecressie}.

In practice, the summation in (\ref{eqn:FFHdisc}) is truncated, resulting in a new model for $\{\theta_i\}$:
\begin{eqnarray}
\theta_{i}=\beta_0+\sum_{k=1}^{K} b(k) \xi_{i}(k)+ \mathbf{x}_{i}'\boldsymbol{\beta}_{x}+u_{i};i=1,\ldots,n,
\label{eqn:Trunc}
\end{eqnarray}
where $K<T$ and, with a slight abuse of notation, $\{u_i\}$ is not give a different symbol. Then equations (\ref{eqn:Main}) and (\ref{eqn:Trunc}) together represent a FH model that includes both scalar and functional covariates. Typically, $K$ is chosen such that some predetermined percentage (e.g., 95\%) of variation in the function is retained.  That is, $K$ is the smallest integer such that $\sum_{k=1}^K\widetilde{\lambda}_{k}/\sum_{k=1}^T\widetilde{\lambda}_{k} \ge 0.95$.  However, in our framework, this only represents an initial phase of dimension reduction.  Subsequent dimension reduction proceeds by stochastic search variable selection (SSVS) \citep{george1993variable, george1997approaches}. Note that, for $J>1$, the truncation number $K$ typically depends on the specific function; that is, $K$ is replaced with $\{K_j\}$.

Bayesian SSVS requires prior distributions for the components of $\mbf{b} \equiv (b(1),\ldots,b(K))'$ and of $\mbv{\beta}_x$ in (\ref{eqn:Trunc}).  In general, when interest resides in a substantial number of submodels, as is the case in the examples we consider, SSVS algorithms provide an effective means of model selection \citep[e.g., see][for a comprehensive overview]{george2000variable}.  Returning to the case where $J\ge 1$ (i.e., $j=1,\ldots,J$), let $\mbf{b}_j\equiv (b_j(1),\ldots,b_j(K))'$. Following \cite{george1993variable}, we use the mixed-normal prior distribution,
\begin{eqnarray}
b_j(k) | \gamma_{jk} \sim \gamma_{jk} N(0,c_{jk} \tau_{jk}) + (1 - \gamma_{jk}) N(0,\tau_{jk}); \;\; k=1,\ldots,K_j,
\label{eq:SSVSprior1}
\end{eqnarray}
where conditional independence of $\{b_j(k)\}$ is assumed, and $\{\gamma_{jk}\}$  are specified at the next level of the hierarchy to have independent Bernoulli($\pi_{jk}$) distributions, with parameter $0 < \pi_{jk} < 1$, for $k=1,\ldots,K_j$.  In this context, $\pi_{jk}$ represents the prior probability that $b_j(k)$ should be included in the model, and $\gamma_{jk}=1$ indicates that the $k$-th expansion coefficient ($k=1,\ldots,K_j$) for the $j$-th functional covariate $(j=1,\ldots, J)$ is included in the model.  Now, typically, $c_{jk}$, $\tau_{jk}$, and $\pi_{jk}$ are taken as fixed hyperparameters; \citet{george1993variable,george1997approaches} present several alternatives for their specification.  Specifically, they recommend taking $\tau_{jk}$ to be small so that, when $\gamma_{jk} = 0$, it is sensible to specify an effective prior for $b_j(k)$ that is close to zero.  Additionally, in general, it is advantageous to take $c_{jk}$ to be large (greater than 1) so that if $\gamma_{jk}= 1$, then the prior favors a nonzero $b_j(k)$.
Selection of the elements of $\mbv{\beta}_x$ proceeds in an identical manner to selection of the elements of a $\mbf{b}_j$.  Joint selection proceeds for $\{\mbf{b}_j: j=1,\ldots, J\}$ and $\mbv{\beta}_x$ by assuming prior mutual independence between all $\{\mbf{b}_j\}$ and $\mbv{\beta}_x$.  When performing SSVS, it is important to standardize the functional components and covariates so that they are on the same scale. Otherwise, certain covariates may be selected frequently based solely on their magnitude. Therefore, in our simulations as well as in our analysis of ACS's percent household Spanish-speaking data, all the functional principal components are scaled to have unit variance.  For further discussion surrounding SSVS as it relates to functional data modeling, see \citet{holan2010modeling,holan2012approach} and the references therein.

The prior described by (\ref{eq:SSVSprior1}) reflects one option among several choices available in the literature on Bayesian variable selection \citep{o2009review, vannucci2010}; we use it because it has been shown to work well in similar settings \citep[e.g., see][]{holan2010modeling,holan2012approach}. SSVS algorithms that assume dependence among the $\{b_j(k)\}$ through hierarchical priors could also be considered \citep[e.g., ][]{yang2013ecological} and may be used to target selection of certain coefficients.

\subsection{Spatial Random Effects}
Most extensions of the basic FH model assume independent Gaussian latent random effects for $\mbf{u}=(u_{1},u_2,\ldots,u_{n})'$. Instead, the model we propose assumes spatially correlated random effects based on the ICAR model, but other spatial models could be used \citep[see, e.g.,][for a review and comparison of these]{senguptacressie2013}. In SAE, the use of the ICAR model dates to back to \cite{besag1991}, \cite{leroux1999estimation} and \cite{macnab2003hierarchical}, who utilize such a model to estimate rates for non-rare diseases in small areas. CAR and ICAR models have also been employed in the FH context \citep[e.g.,][]{cressie1990small,gomez2010bayesian,you2011hierarchical}. In addition, \cite{torabi2011hierarchical} has implemented the ICAR model to account for the spatial effects in a spatio-temporal hierarchical Bayesian FH model.  We utilize the same ICAR structure here, now in the presence of functional covariates.  Our choice of an ICAR structure, over other models of spatial dependence (such as SAR models), is primarily based on its parsimonious specification and its ability to capture relatively smooth spatial dependence.

The use of ICAR random effects allows the latent spatial characteristics of the data to be modeled directly, which facilitates the borrowing of strength across spatial units. The ICAR formulation is due to \cite{besag1991}. In ({\ref{eqn:FFH}), define
\begin{eqnarray}
u_{i}|\{u_{j \neq i}\}\sim N \left(\sum_{i \sim j} \frac{u_{j}}{w_{i+}}, \frac{\sigma_{u}^{2}}{w_{i+}}\right),
\label{eqn:ICAR}
\end{eqnarray}
where the notation ``$i \sim j$" denotes that small areas $i$ and $j$ are neighbors (i.e., they share a border), and $w_{i+}$ is the number of neighbors associated with small area $i$. The ICAR model defined by (\ref{eqn:ICAR}) yields an Intrinsic Gaussian Markov Random Field (IGMRF) \citep{rue2005gaussian}, which corresponds to an improper prior distribution on $\{u_i\}$ in the hierarchical model we propose. Let $\Sigma_u$ denote cov$(u_1,\ldots,u_n)$; then the precision matrix of this IGMRF has the form,
\begin{eqnarray*}
\mbv{\Sigma}_{u}^{-1}=\sigma_{u}^{-2} (\mbf{D}_{w}-\mbf{W}),
\end{eqnarray*}
where $\mbf{D}_{w}$ is a diagonal matrix with element $(i,i)$ equal to $w_{i+}$. Further, the $(i,j)$-th element of $\mbf{W}$ equals one if small areas $i$ and $j$ are neighbors, and it equals zero otherwise. The diagonal of $\mbf{W}$ is set to zero since small area $i$ is not a neighbor of itself.

The improper prior on $\{u_i\}$ is due to a linear dependency in the columns of $(\mbf{D}_{w}-\mbf{W})$, which can be seen by post-multiplying this matrix by a vector of ones and noting that it yields a vector of zeroes. Despite its impropriety, the ICAR prior distribution is often used, as it yields a proper posterior distribution for many commonly used data models, such as the Gaussian, Poisson, and Binomial distributions. The ICAR prior implies a smoother spatial process than can be obtained from a CAR prior, and hence it facilitates more borrowing of strength between spatial units.  A ``sum-to-zero" constraint, $\sum_{i=1}^{n} u_{i}=0$,  is needed to allow the intercept term in the model to be estimable; if not enforced, the intercept and the spatial random effects, $\{u_i\}$, are linearly dependent. Fast algorithms for sampling $\{u_i\}$ subject to $\sum_{i=1}^{n} u_{i}=0$, can be found in \cite{rue2005gaussian} and are used in our simulations and data analysis (Sections~\ref{SecSims} and \ref{SecData}).

As previously noted, in conjunction with a Gaussian data model, the ICAR prior yields a proper Gaussian posterior distribution for $\{u_i: i=1,\ldots,n\}$. This makes the ICAR (and CAR models in general) convenient for modeling the spatial dependency in the FH framework, where Gaussian data models are typically assumed. In a hierarchical modeling framework, of which the FH model is a special case, the posterior distribution can often be sampled using a Markov chain Monte Carlo (MCMC) algorithm known as the Gibbs sampler. When an ICAR or CAR prior is used with a non-Gaussian data model, Bayesian inference typically proceeds using a Metropolis-within-Gibbs MCMC algorithm.

\section{Simulation Study}\label{SecSims}
The simulation study we consider is designed to evaluate the performance of our model (\ref{eqn:Main}), (\ref{eqn:FH}), (\ref{eqn:Trunc}), (\ref{eq:SSVSprior1}), and (\ref{eqn:ICAR}) using simulated data that is calibrated to behave like our motivating example using ACS's percent household Spanish-speaking data. In particular, we consider the effect of using both functional-covariate information and spatial correlation, within the FH context. In this simulation study, we only utilize curves associated with the search term ``y," which were seen, through exploratory methods, to contain significant auxiliary information in predicting the responses $\{\theta_1,\ldots,\theta_n\}$.

Using the expansion coefficients from (\ref{eqn:Trunc}), based on the detrended time series (see Step 2, Appendix A), we generated 250 datasets according to the algorithm given in Appendix A.  For each dataset, we estimated a FH model with an ICAR spatial structure using SSVS.  Our MCMC algorithm consisted of 50,000 iterations with the first 2,000 discarded for burn-in. In this setting, all of the full conditional distributions are of standard form and straightforward to derive (Appendix B). Consequently, Gibbs sampling was used for inference on all model parameters.   The model used for generating the simulated data $Y^*_i$ is, for $i=1,\ldots,n$,
\begin{eqnarray*}
Y^*_{i}&=&\widehat{\theta}_{i}+\epsilon_{i}\\
\widehat{\theta}_{i} &=& \beta_0+\sum_{k=1}^K b(k)\widehat{\xi}_{i}(k)+u_{i},
\end{eqnarray*}
where the superscript ``$*$" distinguishes the real data from the survey estimates analyzed in Section 5, $K=13$, and $\widehat{\xi}_i(k)$ is derived from $\widehat{\mathbf{z}}_{i}(t)-\overline{\mathbf{z}}$, with $\widehat{\mathbf{z}}_{i}(t)$ corresponding to the Google Trends curves for the search term ``y."   Finally, $\{u_i\}$ is assumed to follow the ICAR structure specified in (\ref{eqn:ICAR}) with parameters detailed in Step 5 of Appendix A.

For each of the 250 datasets we fit the model made up of (\ref{eqn:Main}) and the particular case of (\ref{eqn:FFHdisc}) given by
\begin{eqnarray*}
\theta_i=\beta_0+\sum_{k=1}^{13}b(k)\xi_i(k)+u_i.
\label{eq:simmod}
\end{eqnarray*}
In this case, $\{u_i: i=1,\ldots,n\}$ follows the ICAR model given in (\ref{eqn:ICAR}), with $\sigma^2_u\sim IG(0.001,0.001)$ and a ``sum-to-zero" constraint imposed on the elements of $\{u_i\}$.  Finally, we assume $\beta_0\sim  N(0, \sigma_\beta^2)$, with $\sigma_\beta^2\sim IG(0.001,0.001)$.

Our primary interest is in reducing the MSE of the survey quantity of interest, namely the superpopulation mean for area $i$. For each of the 250 simulated datasets, three analyses were performed. The first analysis was performed using the Spatial FH model with functional covariates; see (\ref{eqn:Trunc}) (henceforth called the ``SFFH" model). The second analysis was performed using a FH model with functional covariates and independent Gaussian spatial effects, independent Gaussian effects being typical in the FH framework (henceforth called the ``FFH" model). The third analysis was performed with latent spatial effects but no functional predictors (henceforth called the ``Spatial Only" model).  Prior specifications for the SFFH model are identical to those used in our analysis below of the ACS's percent household Spanish-speaking data (Section~\ref{SecData}). Priors for the functional covariates in the FFH model, and priors for the latent spatial effects of the Spatial Only model are identical to those in the SFFH model. Table~\ref{Ta:Sim} summarizes these results.

As illustrated in Table~\ref{Ta:Sim}, we see that the SFFH model outperforms the other two models in 13 out of 21 locations and provides the lowest overall MSE, $\sum_{i=1}^n(Y_i-\widehat{Y}_i)/n$, making it the preferred model in these simulations. The Spatial Only model performs second best, providing the lowest MSE in seven out of 21 locations and the second lowest MSE overall. In this context, it is clear that the combination of spatial and functional information is preferred over using either type of information alone.

\section{Google Trends Data to Improve ACS Estimates}\label{SecData}
Recall that we utilize a prior distribution for SSVS that consists of a mixture of normals  to distill the important features of the functional covariates. When employing the SSVS procedure, it is typically advantageous to ensure that all of the covariates are on the same scale. Otherwise, certain components may be selected based solely on their relative magnitude. Therefore, in addition to the standardization discussed in Section~\ref{SecMotive}, in our model we standardized the collection of the expansion coefficients, $\{\xi_{ij}(k)$, from Section~\ref{subsec:dimred} to have mean zero and unit variance within each function.

The model we consider differs from the simulation study (Section~\ref{SecSims}) in that we utilize the search terms ``y" and ``yo" as our functional covariates (see Figure~\ref{Fi:Functional}).  The reason for exclusion of the search term ``el" is that, when combined with the other search terms, there are principal-component combinations that completely remove the spatial dependence.  Hence, we want to ``stress test" our model by purposely leaving out covariate information and allowing the spatial component to capture it. For the two covariates we keep (i.e.,  $J=2$ here), we utilize the entire functionals and identify the most important features (using SSVS), devoid of needing to {\it a priori} select user-defined curve features. Our final model for the relative change of percent household Spanish-speaking is, for $i=1,\ldots,21$,
\begin{eqnarray}\label{eq:FHACS}
 Y_{i}&=&\theta_{i}+\epsilon_{i}\\
\nonumber \theta_{i}&=&\beta_0+\sum_{j=1}^{2}\sum_{k=1}^{13} b_{j}(k) \xi_{ij}(k)+ \mathbf{x}_{i}'\boldsymbol{\beta}_{x}+u_{i},
\end{eqnarray}
where $\{\epsilon_i\}$ are independent Gaussian random variables with mean zero and variance $\{\sigma^2_i\}$, respectively, and the remaining terms are defined in (\ref{eqn:Trunc}).  The sampling variance associated with $\epsilon_i$ in (\ref{eq:FHACS}), namely $\sigma^2_i$, for $i=1,\ldots,21$, is obtained using the delta method from variances provided by the U.S Census Bureau based on a Successive Difference Replication (SDR) method \citep{USCB:ACSMan}.  In our context, we consider this variance known, as is common in SAE methodology.

For our purposes, $\pi_{jk}$ in (\ref{eq:SSVSprior1}), which is the SSVS portion of the model, was fixed at 0.5 for $j=1,2$ and for all $k$, as this yields equal contributions to the likelihood whether a variable is included or not, and in this sense it can be considered noninformative. We used the parameterization $c_{jk} \equiv c$ for all $k$ and $\tau_{jk} \equiv \tau$ for $j=1,2$ and for all $k$, with $c$ and $\tau$ chosen via a sensitivity analysis. Specifically, we allowed $\tau$ to take values $10^{-3}$, $10^{-4}$, and $10^{-5}$, and $c$ to take values 10 and 100.  A factorial (sensitivity) experiment was performed in order to select the values of $c$ and $\tau$ for our analysis. In this experiment, we chose the values of $c$ and $\tau$ that yielded the lowest within-sample MSE. For each combination of $c$ and $\tau$, the MCMC algorithm consisted of 50,000 iterations with the first 2,000 iterations discarded as burn-in. The remaining 48,000 iterations for each small area were then used for inference.  Our factorial experiment selected $\tau=10^{-5}$ and $c=10$ as producing the lowest MSE.

Fixing $c=10$ and $\tau=10^{-5}$, we ran a leave-one-out analysis on the ACS data $\{Y_i\}$.  The MCMC algorithm for each location consisted of 50,000 iterations with the first 2,000 iterations being discarded for burn-in.  The posterior mean of the predicted value at each left-out location, $\widehat{Y}_{-i}$, and the empirical mean squared prediction error (MSPE) across all locations, namely $ \sum_{i=1}^n(Y_i-\widehat{Y}_{-i})/n$, where $n=21$, were computed. The leave-one-out MSPE for the SFFH model is $3.78\times 10^{-3}$; for the FFH model the leave-one-out MSPE is $5.17 \times 10^{-3}$, and for the Spatial Only model the leave-one-out MSPE is $3.85 \times 10^{-3}$.  The MCMC algorithm consisted of enough iterations to verify that these differences are not due to Monte Carlo error.  This analysis illustrates that the SFFH model is preferred to the other two models in terms of leave-one-out MSPE. The individual squared deviations, $(Y_i-\widehat{Y}_{-i})^2$, are provided in Table~\ref{Ta:Data}. The FFH model was left out of the table due to its inferior overall MSPE.

From the table, we see that the Spatial Only model yields estimates closer to the observations in 11 of 21 locations.  However, when the Spatial Only estimates are closer to the observed values they are not substantially closer.  In contrast, when the SFFH model provides closer estimates, they are frequently far superior to the Spatial Only model. These results occur because, in several locations, the posterior distribution of the spatial process places the majority of its mass near zero, indicating that the functional covariates are accounting for spatial dependence.  That is, the SFFH estimates in these locations tend to be quite similar to the Spatial Only estimates, but with slightly more variation, which contributes to slightly inferior estimation.  However, in several locations, the posterior mass of the spatial latent effects is far from zero and, in these locations, the SFFH model provides superior estimates.

It is often of interest to examine the first several functional principal components, as well as the components selected most frequently within the SSVS.  The former (Figure~\ref{Fi:firstcomp}) captures the features accounting for the majority of the variation, whereas the latter (Figure~\ref{Fi:chosencomp}) illustrates which functional aspects feature most heavily in estimating the superpopulation mean.  Figure~\ref{Fi:chosencomp} shows that high-frequency principal components play an important role in determining the functional covariates.

In Table~\ref{Ta:Data} and Figure~\ref{Fi:Data}, we provide the ratio of the squared deviation $(Y_i-\widehat{Y}_{-i})^2$ for the SFFH, divided by the squared deviation for the Spatial Only model, when the Spatial Only model provides a better estimate; and we provide its inverse when the SFFH model provides a better estimate. When the Spatial Only model performs better, the mean ratio is 4.50, whereas the mean ratio is 29.95 when the SFFH model performs better.   We note that this value is greatly affected by the SFFH model's far superior performance in predicting Wisconsin, which provides a ratio of 230.48.  Despite its substantial influence, Wisconsin does not drive the results above; the mean ratio is still 6.68 with Wisconsin removed from consideration. We conclude that the SFFH model is beneficial in modeling these data when the functional covariates alone do not completely account for spatial dependence, and we also conclude that the SFFH model provides lower overall leave-one-out MSPE.

Finally, a natural question that arises is how our estimates compare with those of the public-use ACS data.  For the survey itself, we cannot compute a model-based leave-one-out estimate.  Additionally, because we do not have the true values, we cannot compute the MSE  similar to the simulation and the cross-validation.  However, we do have the ability to compare the precision of our model-based estimates relative to those of the public use data.  Figure~\ref{Fi:ratios} provides a plot of the log standard deviations of the model-based estimates $\{\widehat{\theta}_i\}$ of the SFFH model versus the log sampling standard errors provided by the ACS.  This figure demonstrates that the SFFH model provides more precise estimates in all 21 locations, with higher relative precision in those areas which the ACS estimates have higher sampling standard errors. In fact, letting $\mbox{var}(\widehat{\theta}_{i_{\tiny{ACS}}})$ denote the sampling variance of the ACS and $\mbox{var}(\widehat{\theta}_{i_{\tiny{SFFH}}})$ denote the model-based variance of $\theta_i$, the mean relative reduction in variance is 21\%, where the relative reduction in variance for each location is given by $\{\mbox{var}(\widehat{\theta}_{i_{{\tiny ACS}}})-\hbox{var}(\widehat{\theta}_{i_{\tiny SFFH}})\}/\mbox{var}(\widehat{\theta}_{i_{\tiny ACS}})$.


\section{Discussion}\label{SecDis}
Fay-Herriot models have a celebrated history, owing to their versatility in small area estimation.  To increase the usefulness of this class of models, we have extended them to include functional covariates along with spatial dependence.  Importantly, we have demonstrated that dimension-reduced functional covariates can be effectively utilized to improve estimation in public-use ACS data. Further, we have emphasized the importance of the spatial relationships between small areas in our model.

Our fully Bayesian procedure incorporating dimension-reducing SSVS provides an automated method for feature selection and selection among different candidate models. The model selection is tuned to minimize the MSE of $\{\theta_i$: $i=1,\ldots,n\}$, where recall that the MSE is the expected posterior variance. However, it would also be possible to consider other posterior distributional properties, when selecting SSVS hyperparameters.

The issue of spatial dependence has been addressed systematically, and we have illustrated, via model-based simulation and through the ACS's data on percent household Spanish-speaking, that models with spatial autocorrelation yield lower MSEs than non-spatial models. We note that, for these data, the SFFH model, using Google Trends data for the search terms ``y" and ``yo,"  consistently outperforms the FFH model, and this points to the importance of explicitly accounting for spatial association even at geographies as coarse as the state level. We also note that, with ``Big Data" functional covariates, it is possible to collect enough covariates that one may account for the spatial structure in the data (as with the inclusion of all three search terms in the model) and that the SSVS prior facilitates the selection of covariates to achieve dimension reduction.

Due to data limitations of Google Trends, we have applied our approach at the state level, but not for smaller geographies. Twitter data are another source of functional covariates, and they are available at finer spatial resolutions. However, the drawback of using Twitter data is that they are not as readily available.  Finally, our model is also generally applicable to image data, such as remotely sensed scenes of land-use/land-cover, indicating a key potential use of this technique in agricultural surveys.

\section*{Acknowledgments}
This research was partially supported by the U.S. National Science Foundation (NSF) and the U.S. Census Bureau under NSF grant SES-1132031, funded through the NSF-Census Research Network (NCRN) program. We would like the thank the referees and editor for their comments, which have resulted in an improved paper.

\section*{Appendix A: The Simulation Algorithm}
\begin{appendix}
\Appendix    
\renewcommand{\theequation}{A.\arabic{equation}}
\setcounter{equation}{0}

The following algorithm was used to generate the functional covariates and the data for the simulation study presented in Section~\ref{SecSims}.
\begin{description}
\item[Step 1:]  Consider the Google Trends time series for the search term ``y" at location $i$. Denote this quantity by $\mbf{z}_i=(z_i (t_1),...,z_i(t_T))'$.  Let $\mbf{Z}=[\mbf{z}_1,...,\mbf{z}_{n}]$ be a $T \times n$ matrix containing the Google trends time series associated with the search term ``el."
\item[Step 2:] Subtract the location-averaged temporal mean of the matrix $\mbf{Z}$, namely $\bar{\mbf{z}} \equiv n^{-1}(\sum_{i=1}^{n} \mbf{z}_{i})$, from each column of $\mbf{Z}$ to obtain $\mbf{Z^*}$, a matrix of detrended time series.
\item[Step 3:] Consider the $T \times T$ empirical covariance matrix $\mbf{S}^*\equiv \mbf{Z}^* \mbf{Z}^{*'}/(n-1)$. Let $\mbf{S}^*=\mbv{\Phi}^* \mbv{\Lambda}^* \mbv{\Phi}^{*'}$ be the usual spectral decomposition of $\mbf{S}^*$. Here, $\mbv{\Phi}^*$ represents the discretized eigenfunctions for the functional covariate ``el."
\item[Step 4:] Analyze the original ACS data using the discretized eigenfunctions $\mbv{\Phi}^*$ coming from the SFFH model, in order to obtain posetrior-mean values $\widehat{\beta}_0$, $\widehat{\mbf{b}}$, and $\widehat{\sigma}_u^2$, obtained from the posterior distributions of the model parameters. The terms $\{b(k)\}$ used in the simulations are the posterior-mean values of the corresponding parameters obtained from the analysis of the ACS data presented in Section~\ref{SecData}.  This analysis resulted in 26 ``$b(k)$" parameters (corresponding to 13 for ``y" and 13 for ``yo"). It is worth noting here that, in computing the posterior means, every realization of $b(k)$ was utilized, regardless of whether $\gamma_{jk}$ was 0 or 1. This results in our using model-averaged $b(k)$ values in the simulation, ensuring that the larger values of $b(k)$ correspond to important functional principal components.
\item[Step 5:] Simulate a set of responses, $\mbf{Y}^*=\widehat{\beta}+\mbv{\Phi}^{*} \widehat{\mbf{b}} +\mbf{u}+ \mbv{\epsilon}$, where $\mbf{u}$ is distributed as a zero-mean ICAR process with parameters obtained by plugging in the estimates from the data analysis in Section~\ref{SecData}; $\mbv{\epsilon} \sim MVN \left(\mbf{0},\hbox{diag}(\sigma_1^2,\ldots,\sigma_n^2)\right)$, where $\sigma^2_i$, $i=1,\ldots,n$, are the known survey variances (see Section~\ref{SecTheory}). $\mbf{Y}$ denotes the $n$-dimensional vector of observed small-area responses from the ACS, namely (\ref{eqn:Outcome}).
\end{description}

\end{appendix}

\section*{Appendix B: Full Conditional Distributions}

\begin{appendix}
\Appendix    
\renewcommand{\theequation}{B.\arabic{equation}}
\setcounter{equation}{0}

Here we provide the forms of the full conditional distributions for the SFFH model utilized in Section~\ref{SecData}. We define $\mbv{\Upsilon}$ as a block diagonal matrix with diagonal entries equal to $c \tau \gamma_{jk} + \tau(1-\gamma_{jk})$; $j=1,\dots,J$, $k=1,\ldots, K_j$, and we define $\mbv{\Sigma}_{{\epsilon}}$ to be $n \times n$ diagonal matrix with $\Sigma_{{\epsilon},ii}=\sigma_{i}^{2}$. The $n\times K_J$ matrix, $\mbv{\Xi}=[\mbv{\xi}_{1}(1),\ldots,\mbv{\xi}_{J}(K_J)]$, has columns $\mbv{\xi}_j(k)=(\xi_{1j}(k),\ldots,\xi_{nj}(k))'$, and recall that $n$ is the number of small areas under consideration. For our analysis, $n=21$, and we let $\mbf{b}=(\mbf{b}_1',\ldots,\mbf{b}_J')'$ denote the concatenated $K_+$-dimensional vector of $\{\mbf{b}_j\}$, where $K_+=\sum_j K_j=26$ for our analysis. The scalars $a_1$ and $a_2$ denote the shape and scale parameters in the $IG(a_1,a_2)$ prior for $\sigma_{{u}}^{2}$ and $\sigma_{{\beta}}^{2}$. For our analysis, we set $a_1=a_2=0.001$.  Under this notation, the full conditional distributions  have the following forms:
\begin{enumerate}
\item $\mbf{b}|\mbf{u},\{\gamma_{jk}\},\sigma_u^2,\beta_0,\sigma_{\beta_0}^{2},\mbf{Y} \sim MVN(\mbv{\mu}_{b}, \mbv{\Sigma}_{b})$, where\\
$\mbv{\Sigma}_{b}=(\mbv{\Xi}'\mbv{\Sigma}_{\epsilon}^{-1} \mbv{\Xi}+\mbv{\Upsilon}^{-1})^{-1}$ and \\
$\mbv{\mu}_{b}=\mbv{\Sigma}_{b} \mbv{\Xi}' \mbv{\Sigma}_{\epsilon}^{-1}(\mbf{Y}-\mbf{1} \beta_0 -\mbf{u})$.
\item $\mbf{u}|\mbf{b},\{\gamma_{jk}\},\sigma_u^2,\beta_0,\sigma_{\beta_0}^{2},\mbf{Y} \sim MVN(\mbv{\mu}_u, \mbv{\Omega}_{u})I_{\{\sum_{i=1}^n u_i=0\}}$, where\\
$\mbv{\Omega}_{u}=(\mbv{\Sigma}_{\epsilon}^{-1}+\sigma_{u}^{-1}\{\mbf{D}_{w}-\mbf{W}\})^{-1}$,\\
 $\mbv{\mu}_{{u}}=\mbv{\Omega}_{{u}} \mbv{\Sigma}_{\epsilon}^{-1}(\mbf{Y}-\mbf{1} \beta_0 -\mbv{\Xi} \mbf{b})$, and $I_{\{\cdot\}}$ denotes the indicator function.
\item For $k=1,\ldots,K_j$, and $j=1,\ldots,J,$
\begin{eqnarray*}
\gamma_{jk}|\mbf{b},\mbf{u},\{\gamma_{-jk}\},\sigma_u^2,\beta_0,\sigma_{\beta_0}^{2},\mbf{Y} \sim Bern\left(\frac{f(b_{jk}|\gamma_{jk}=1)}{f(b_{jk}|\gamma_{jk}=1)+f(b_{jk}|\gamma_{jk}=0)}\right),
\end{eqnarray*}
where $f(\cdot)$ is the pdf of the normal prior associated with $b_{jk}$, and $Bern(p)$ denotes a Bernoulli distribution with probability $p$.
\item $\sigma_{u}^{2}|\mbf{b},\mbf{u},\{\gamma_{jk}\},\beta_0,\sigma_{\beta_0}^{2},\mbf{Y} \sim IG(a_1+ n/2, a_2 + \mbf{u}'(\mbf{D}_{w}-\mbf{W})\mbf{u}/2)$.
\item $\beta_0|\mbf{b},\mbf{u},\{\gamma_{jk}\},\sigma_u^2,\sigma_{\beta_0}^{2},\mbf{Y} \sim N(\mu_{\beta_0}, \widetilde{\sigma}_{\beta_0}^2)$, where\\
$\widetilde{\sigma}_{\beta_0}^2=(\mbf{1}' \mbv{\Sigma}_{\epsilon}^{-1} \mbf{1}+{\sigma_{\beta_0}^{2}}^{-1})^{-1}$ and\\
 $\mu_{\beta_0}=\widetilde{\sigma}_{\beta_0}^2\mbf{1}' \mbv{\Sigma}_\epsilon^{-1}(\mbf{y}-\mbv{\Xi}\mbf{b}-\mbf{u})$.
\item $\sigma_{\beta_0}^{2}|\mbf{b},\mbf{u},\{\gamma_{jk}\},\sigma_u^2,\beta_0,\mbf{Y} \sim IG(a_1+1/2, a_2 + \beta_{0}^2/2)$.
\end{enumerate}
Finally, although we did not include any scalar covariates, they can be handled straightforwardly.  That is, sampling $\mbv{\beta}_x$ in (\ref{eqn:FFH}) using an SSVS prior would proceed in a similar manner to sampling the functional covariates \citep[see][for an example]{holan2012approach}.

\end{appendix}

\clearpage\pagebreak\newpage

\baselineskip=14pt \vskip 4mm\noindent
\bibliographystyle{jasa}
\bibliography{STSN}

\clearpage\pagebreak\newpage
\clearpage

\begin{table}\footnotesize
\begin{center}
\begin{tabular} {|c|c|c|c|} \hline
State & SFFH&FFH & Spatial Only \\ \hline
Alabama&3.27 & \textbf{2.60} & 4.47\\
Connecticut& 0.57 & 1.63 & \textbf{0.55}\\
District of Columbia & \textbf{ 4.45} & 4.53 & 5.21\\
Florida & \textbf{ 0.06} & 0.86 & 0.06\\
Georgia & 0.30 &1.17 & \textbf{0.29}\\
Illinois  &0.13 & 1.23 & \textbf{0.13}\\
Indiana  & \textbf{0.91} & 1.73 & 1.03\\
Kentucky &  1.64 & 1.89 & \textbf{1.41}\\
Maryland  &\textbf{1.08} & 1.72 & 1.10\\
Massachusetts &  0.53 & 1.71 & \textbf{0.53}\\
Michigan &\textbf{1.58} & 1.59 & 1.91\\
Minnesota &  \textbf{1.58} & 2.55 & 1.74 \\
Missouri  &  \textbf{1.48} &  1.91 & 1.57\\
New Jersey  &  \textbf{0.21} & 0.48 &  0.22\\
New York  & 0.08 & 0.54 & \textbf{0.08}\\
North Carolina  &0.38 & 0.57 &\textbf{ 0.36}\\
Ohio  &\textbf{ 0.80} & 1.91 & 0.85\\
Pennsylvania  & \textbf{0.52} & 0.69 & 0.62\\
South Carolina &  \textbf{1.54} & 3.31 &  1.81\\
Tennessee  & \textbf{1.20} & 1.79 & 1.52\\
Wisconsin  & \textbf{0.86} & 1.41 & 0.91\\ \hline
MSE &  \textbf{1.10} & 1.67 & 1.26\\
\hline
\end{tabular}
\end{center}
\caption{\baselineskip=10pt MSE $\times$ 1000 for the 21 small areas based on 250 simulated datasets (Section 4) for the spatial FH model with functional covariates (SFFH), the  FH model with functional covariates (FFH), and the FH model with only spatial random effects (Spatial Only). Bolded values indicate the smallest MSE for each area. For Florida, the SFFH value is  $6.023 \times 10^{-5}$ and the Spatial Only value is $6.065 \times 10^{-5}$. For New York, the SFFH value is $7.894 \times 10^{-5}$ and the Spatial Only value is $7.840 \times 10^{-5}$.}
\label{Ta:Sim}
\end{table}

\begin{table}\footnotesize
\begin{center}
\begin{tabular} {|c|c|c|c|} \hline
State & SFFH& Spatial Only & Ratio \\ \hline
Alabama&\textbf{21.55} & 22.77 &1.06\\
Connecticut& \textbf{0.06} & 0.53 & 8.72\\
District of Columbia &  0.35 & \textbf{0.02} & 17.05\\
Florida &  \textbf{0.29} & 4.12 &14.37\\
Georgia & 5.06 & \textbf{3.16} &1.60\\
Illinois  &0.13 & \textbf{0.07} &2.00\\
Indiana  & \textbf{0.32} & 3.38 &10.45\\
Kentucky &  1.54& \textbf{1.40} &1.10\\
Maryland  &1.55 & \textbf{0.57} &2.72\\
Massachusetts &  \textbf{0.15} & 1.47 & 10.10\\
Michigan &23.90 & \textbf{22.02} &1.08\\
Minnesota & \textbf{ 1.80 }& 2.92 &1.62\\
Missouri  &  9.04 & \textbf{7.71} &1.17\\
New Jersey  &  \textbf{0.02} & 0.16& 10.47\\
New York  & \textbf{0.15} &  0.63 &4.07\\
North Carolina & 2.66 &\textbf{0.24} &10.99\\
Ohio  & 1.94 & \textbf{0.24} &9.41\\
Pennsylvania  & 0.77 &\textbf{0.74}& 1.05\\
South Carolina &  5.56 &\textbf{4.27} & 1.30\\
Tennessee  &\textbf{ 2.57} & 3.09 &1.20\\
Wisconsin  & \textbf{0.01} & 1.39 &230.48\\ \hline
\hline
\end{tabular}
\end{center}
\caption{\baselineskip=10pt Squared leave-one-out deviations, $(Y_i-\widehat{Y}_{-i})^2$ $\times$ 1000, for the 21 small areas for the analysis of the relative changes of percent household Spanish-speaking in the eastern half of the United States (Section~\ref{SecData}). Bolded values indicate the lowest squared deviation and, hence, the preferred model for the location. The Ratio column is the ratio of the larger squared deviation and the smaller squared deviation.}
\label{Ta:Data}
\end{table}

\clearpage\pagebreak\newpage

\begin{figure}
\caption{\baselineskip=10pt Functional covariates (temporal curves) for the Google Trends search loads of ``el," ``yo," and ``y" (see Section~\ref{SecMotive}). To avoid clutter, we show only the first five time series, in alphabetical order (i.e., Alabama, Connecticut, District of Columbia, Florida, and Georgia), for each search term.}
\label{Fi:Functional}
\centerline{\includegraphics[width=100mm, height=170mm,angle=-90]{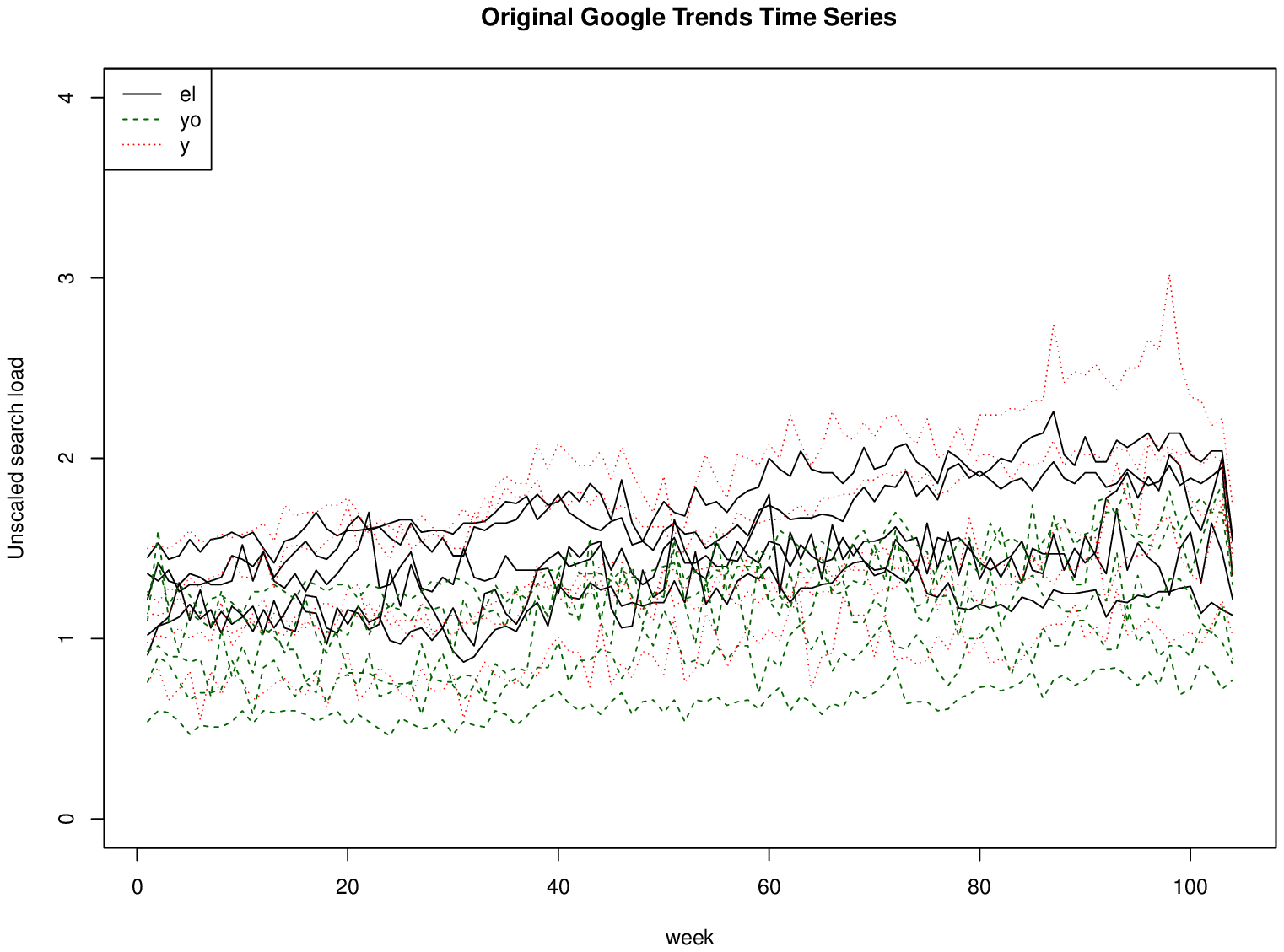}}
\end{figure}

\clearpage\pagebreak\newpage

\begin{figure}[ht]
\caption{\baselineskip=10pt The first four functional principle components of each Google Trends search term ``y" and ``yo" from the data analysis in Section~\ref{SecData}. The search term and component number are listed on the vertical axis. Note that the percentage of variation accounted for by the first four principal components is 95\% for ``y" and 77\% for ``yo."}
\label{Fi:firstcomp}
\centerline{\includegraphics[width=7in, height=7in,angle=-90]{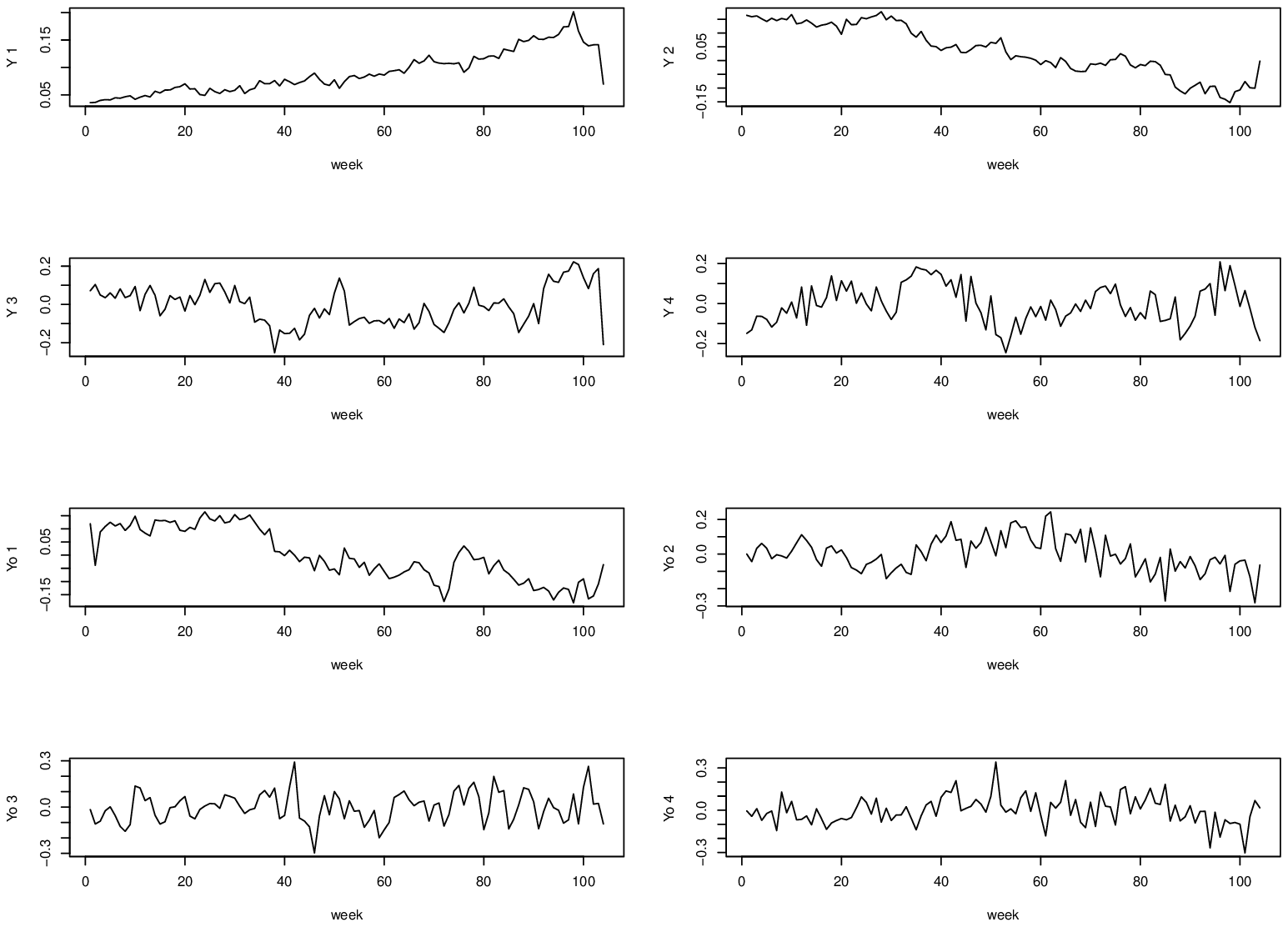}}
\end{figure}

\clearpage\pagebreak\newpage

\begin{figure}[ht]
\caption{\baselineskip=10pt The functional principle components, associated with the Google Trends search term ``y" and ``yo,"  chosen in over 50\% of MCMC iterations of the data analysis in Section~\ref{SecData}.  The search term and component number are listed on the vertical axis. Note that there were five components chosen from ``y" and three components chosen from ``yo."}
\label{Fi:chosencomp}
\centerline{\includegraphics[width=7in, height=7in,angle=-90]{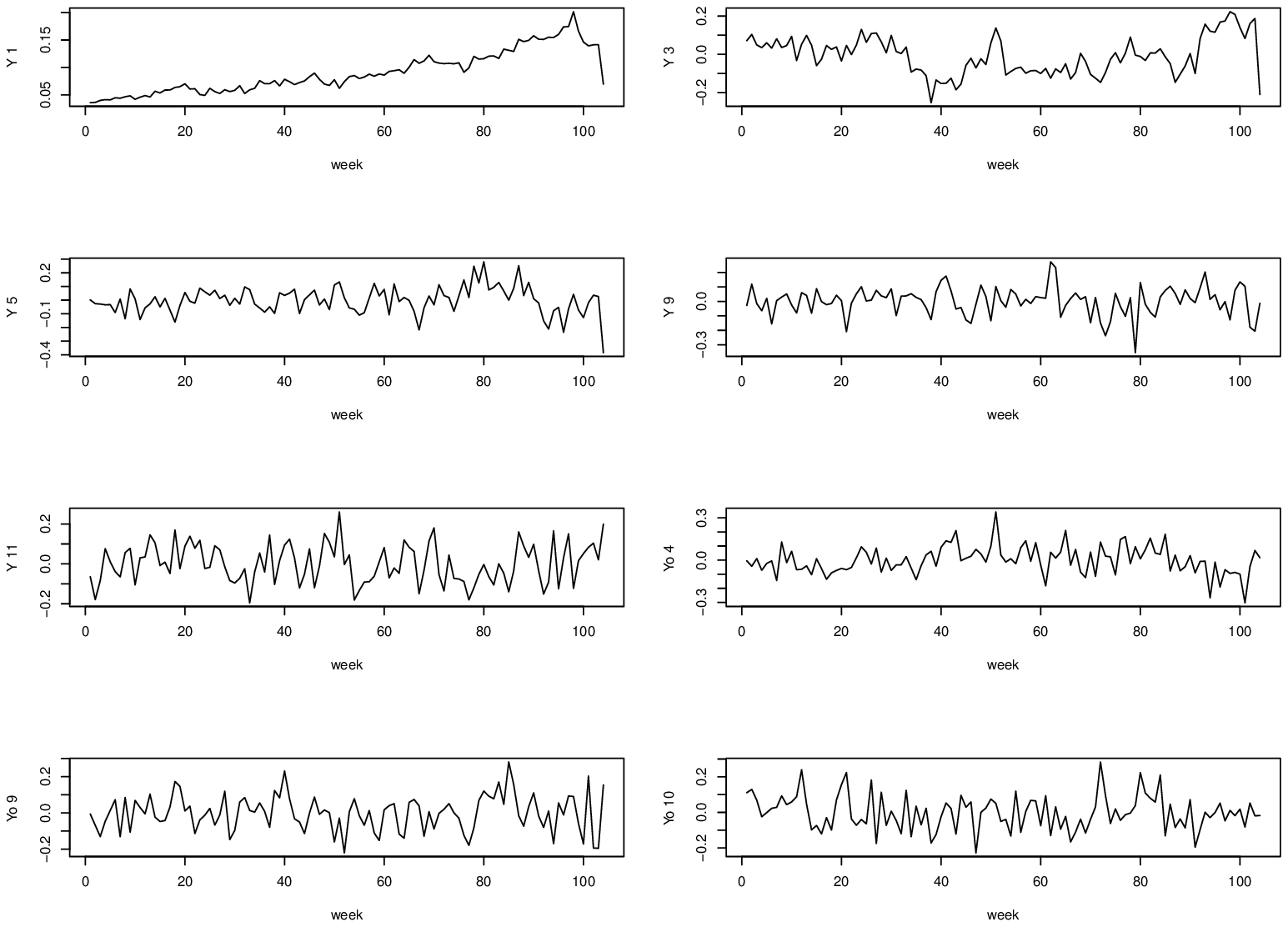}}
\end{figure}


\begin{figure}[ht]
\caption{\baselineskip=10pt  Ratios of the larger to smaller squared deviations for state $i$ for the SFFH model and the Spatial Only model, $i=1,\ldots,21$. Purple indicates areas where the SFFH model is preferred and orange indicates areas where the Spatial Only model is preferred.}
\label{Fi:Data}
\includegraphics[width=140mm, height=180mm,angle=-90]{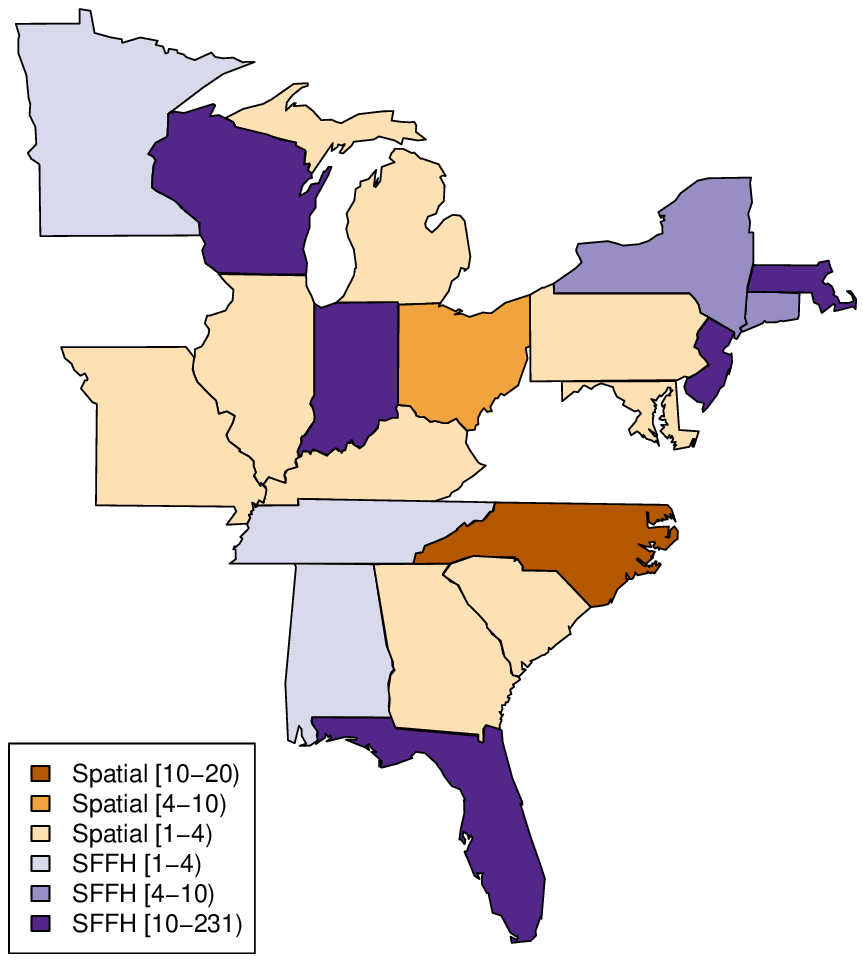}
\end{figure}

\begin{figure}[h]
\caption{\baselineskip=10pt  The log standard deviations of the SFFH model-based estimates of $\{\theta_i\}$ versus the log sampling standard errors of the survey.  The plot demonstrates that the SFFH model always has lower standard deviations than those of the survey estimates, with greater improvement in those areas where the survey yields large sample standard deviation.}
\label{Fi:ratios}
\centerline{\includegraphics[width=100mm, height=170mm,angle=-90]{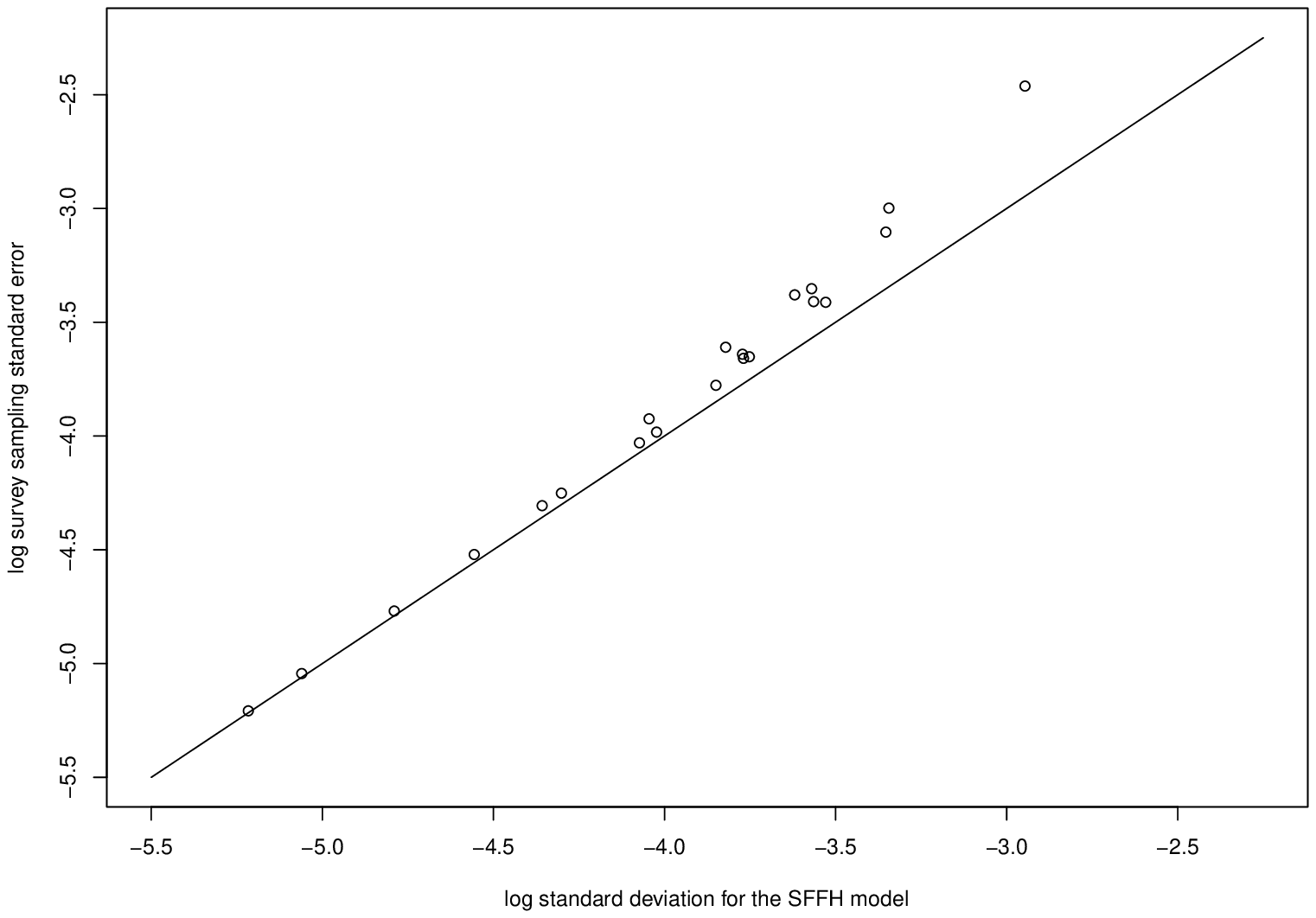}}
\end{figure}

\clearpage\pagebreak\newpage

\end{document}